\documentstyle[12pt,epsfig]{article}
\newcommand{\beq}{\begin{equation}}
\newcommand{\eeq}{\end{equation}}
\newcommand{\beqa}{\begin{eqnarray}}
\newcommand{\eeqa}{\end{eqnarray}}
\newcommand{\ba}{\begin{array}}
\newcommand{\ea}{\end{array}}
\newcommand{\no}{\nonumber}
\newcommand{\ra}{\rightarrow}

\newcommand{\dg}{\dagger}

\newcommand{\wh}{\widehat}
\newcommand{\cL}{{\cal L}}
\newcommand{\cO}{{\cal O}}
\newcommand{\Q}{{\cal Q}}

\newcommand{\dfrac}{\displaystyle \frac}

\newcommand{\nn}{\nonumber \\}
\newcommand{\bea}{\begin{eqnarray}}
\newcommand{\eea}{\end{eqnarray}}
\newcommand{\ol}{\overline}
\normalsize
\begin{document}
\bibliographystyle{plain}
\begin{titlepage}
\begin{flushright}
UWThPh-1999-46\\
FTUV/99-84\\ 
IFIC/99-88\\
\end{flushright}
\vspace{1.5cm}
\begin{center}
{\Large \bf Chiral Perturbation Theory with \\  Photons and
Leptons* $^{\dag}$}
\\[40pt] 

H. Neufeld

\vspace{1cm}
 
Institut f\"ur Theoretische Physik, Universit\"at 
Wien\\ Boltzmanngasse 5, A-1090 Wien, Austria \\[10pt] 

and \\[10pt]
 
Departament de F\'{\i}sica Te\`orica, IFIC, Universitat de 
Val\`encia - CSIC\\ 
Apt. Correus 2085, E-46071 Val\`encia, Spain 
 
\vfill
{\bf Abstract} \\
\end{center}
\noindent
I discuss a low-energy effective field theory which permits the
full treatment of isospin-breaking effects in semileptonic weak
interactions. In addition to the pseudoscalars and the photon, 
also the light leptons have to be included as dynamical degrees of
freedom in an appropriate chiral Lagrangian. 
I describe the construction of the local action at
next-to-leading order.
\vfill
\noindent * Contribution to the Eighth International Symposium
on Meson--Nucleon Physics and the Structure of the Nucleon,
Zuoz, Engadine, Switzerland, August 15-21, 1999.\\
\dag Supported in part by  TMR, EC-Contract No. ERBFMRX-CT980169 
(EURODA$\Phi$NE) and by DGESIC (Spain) under grant No. PB97-1261. 

\end{titlepage}
\addtocounter{page}{1}

\paragraph*{1.}

In this talk, I would like to report about the status \cite{KNRT99} of a
research project on isospin-violating effects in the semileptonic decays
of pions and kaons which is presently carried out by Marc Knecht, Heinz
Rupertsberger, Pere Talavera and myself.  

While isospin-breaking  generated by a non-vanishing quark mass difference 
$m_d - m_u$ is fully contained in the pure QCD sector of the effective
chiral Lagrangian \cite{GL85}, the analysis of isospin-violation of
electromagnetic
origin requires an extension of the usual low-energy effective theory. 
For purely pseudoscalar processes, the suitable theoretical framework   
for the three-flavour case has been worked out in \cite{Urech,NR95,ELM}
by including virtual photons and the appropriate local terms up to 
${\cal O} (e^2 p^2)$. 

The treatment of electromagnetic corrections in 
semileptonic decays demands still a further extension of chiral
perturbation theory. In this case, also the light leptons have
to be included as explicit dynamical degrees of freedom. Only
within such a framework, one will have full control over all
possible isospin breaking effects in the analysis of new high statistics
$K_{\ell 4}$ experiments  by the E865 and KLOE collaborations at
BNL \cite{E865} and DA$\Phi$NE \cite{DAPHNE}, respectively.
The same refined methods are, of course, also necessary for the
interpretation of forthcoming high precision experiments on
other semileptonic decays like $K_{\ell 3}$, etc.

\paragraph*{2.}

To lowest order in the chiral expansion, the
effective Lagrangian without dynamical photons and leptons (pure
QCD) is nothing else than the non-linear sigma model in the presence  
of external vector, axial-vector, scalar and pseudoscalar sources
$v_\mu$, $a_\mu$, $\chi = s +i p$. Following the notation of
\cite{EGPR89}, it takes the form 
\beqa \label{pureQCD} 
\cL_{\rm eff} &=&\frac{F^2}{4} \; \langle u_\mu u^\mu + \chi_+\rangle , 
\eeqa 
where
\beqa 
u_\mu &=& i [u^\dg (\partial_\mu - i r_\mu)u - u
(\partial_\mu - i l_\mu)u^\dg], \no \\ 
l_\mu &=& v_\mu - a_\mu , \no \\
r_\mu &=& v_\mu + a_\mu, \no \\
\chi_+ &=& u^\dg \chi u^\dg + u \chi^\dg u.  
\eeqa
Even this Lagrangian allows the treatment of electromagnetic or
semileptonic processes as long as the photon or the leptons 
are occurring only as external fields. One simply takes external
sources with the quantum numbers of the photon or the $W^{\pm}$.

For the description of dynamical photons and leptons, the extension of 
the lowest order Lagrangian (\ref{pureQCD}) is rather easy.
First of all, the photon field $A_\mu$ and the light leptons 
$\ell,\nu_\ell$ ($\ell = e,\mu$) are introduced in $u_\mu$ by 
adding appropriate terms to the external vector and axial-vector
sources:
\beqa \label{sources}
l_\mu &=& v_\mu - a_\mu - e Q_L^{\rm em} A_\mu + \sum_\ell
(\bar \ell \gamma_\mu \nu_{\ell L} Q_L^{\rm w} + \ol{\nu_{\ell L}} 
\gamma_\mu \ell
Q_L^{{\rm w}\dg}), \no \\
r_\mu &=& v_\mu + a_\mu - e Q_R^{\rm em} A_\mu.
\eeqa   
The $3 \times 3$ matrices $Q_{L,R}^{\rm em}$, $Q_L^{\rm w}$ are additional 
spurion fields. At the end, one identifies $Q_{L,R}^{\rm em}$ with the 
quark charge matrix
\beq \label{Qem}
Q^{\rm em} = \left[ \ba{ccc} 2/3 & 0 & 0 \\ 0 & -1/3 & 0 \\ 0 & 0 & -1/3 \ea
\right],
\eeq
whereas the weak spurion is taken at
\beq \label{Qw}
Q_L^{\rm w} = - 2 \sqrt{2}\; G_F \left[ \ba{ccc}
0 & V_{ud} & V_{us} \\ 0 & 0 & 0 \\ 0 & 0 & 0 \ea \right],
\eeq
where $G_F$ is the Fermi coupling constant and $V_{ud}$, $V_{us}$ are
Kobayashi--Maskawa matrix elements.

Then we have to introduce kinetic terms for the photon and the
leptons and also an electromagnetic term of $\cO(e^2 p^0)$.
With these building blocks, our lowest order effective Lagrangian takes
the form
\beqa \label{Leff}
\cL_{\rm eff} &=& \frac{F^2}{4} \; \langle u_\mu u^\mu + \chi_+\rangle +
e^2 F^4 Z \langle \Q_L^{\rm em} \Q_R^{\rm em}\rangle \no \\
&& \mbox{} - \frac{1}{4} F_{\mu\nu} F^{\mu\nu} + \sum_\ell
[ \bar \ell (i \! \not\!\partial + e \! \not\!\!A - m_\ell)\ell +
\ol{\nu_{\ell L}} \, i \! \not\!\partial \nu_{\ell L}],
\eeqa
where
\beq \label{Qhom}
\Q_L^{\rm em,w} := u Q_L^{\rm em,w} u^\dg, \qquad
\Q_R^{\rm em} := u^\dg Q_R^{\rm em} u.
\eeq

Finally, we have to define an extended chiral expansion scheme. The electric 
charge $e$, the lepton masses $m_e, m_\mu$ and
fermion bilinears are considered (formally) as quantities of order $p$ in the 
chiral counting, where $p$ is a typical meson momentum.
Note, however, that terms of ${\cal O} (e^4)$ will be neglected throughout.

\paragraph*{3.}

As we are dealing with a so-called
non-renormalizable theory, new local terms are arising at
the next-to-leading-order. The associated coupling constants
absorb the divergences of the one-loop graphs. Their finite
parts are in principle certain functions of the parameters of the
standard model. Because of our limited ability in solving the
standard model (confinement problem), these low-energy constants
have to be regarded as free parameters of our effective theory
for the time being.

The list of local counterterms of our extended theory
comprises, of course, the well-known Gasser--Leutwyler
Lagrangian of $\cO(p^4)$ \cite{GL85} and the Urech Lagrangian of 
$\cO(e^2 p^2)$ \cite{Urech} with the generalized $l_\mu$ and
$r_\mu$ defined in Eq. (\ref{sources}). In the presence of virtual
leptons, we have to introduce an additional ``leptonic''
Lagrangian \cite{KNRT99}
\beqa \label{Llept}
\cL_{\rm lept} &=& 
e^2 \sum_{\ell} \left \{ F^2 \left[  
X_1 \ol{\ell} \gamma_\mu \nu_{\ell L} 
\langle u^\mu  \{ \Q_R^{\rm em}, \Q_L^{\rm w} \} \rangle 
\right. \right. \nn [-5pt]
&& \qquad \qquad \left. \left.
+ X_2 \ol{\ell} \gamma_\mu \nu_{\ell L} 
\langle u^\mu  [\Q_R^{\rm em}, \Q_L^{\rm w}] \rangle
\right. \right. \nn
&& \qquad \qquad \left. \left.
+ X_3 m_\ell \ol{\ell} \nu_{\ell L} \langle \Q_L^{\rm w} \Q_R^{\rm em} \rangle
\right. \right. \nn
&& \qquad \qquad \left. \left.
+ i X_4 \ol{\ell} \gamma_\mu \nu_{\ell L} 
\langle \Q_L^{\rm w} \wh \nabla^\mu  \Q_L^{\rm em} \rangle
\right. \right. \nn
&& \qquad \qquad \left. \left.
+ i X_5 \ol{\ell} \gamma_\mu \nu_{\ell L} 
\langle \Q_L^{\rm w} \wh \nabla^\mu  \Q_R^{\rm em} \rangle 
+ h.c. \right]  \right. \nn
&& \qquad  \quad \left.
+ X_6 \bar \ell (i \! \not\!\partial + e \! \not\!\!A )\ell
\right. \nn
&& \qquad  \quad \left.
+ X_7 m_\ell \ol \ell  \ell \right \}. 
\eeqa
where
\beqa \label{covder1}
\wh \nabla_\mu \Q^{\rm em}_L &=& \nabla_\mu \Q^{\rm em}_L 
+ \frac{i}{2} [u_\mu,\Q^{\rm em}_L] =
u (D_\mu Q_L^{\rm em}) u^\dg , \no \\
\wh \nabla_\mu \Q^{\rm em}_R &=& \nabla_\mu \Q^{\rm em}_R 
- \frac{i}{2} [u_\mu,\Q^{\rm em}_R] =
u^\dg (D_\mu Q^{\rm em}_R) u ,
\eeqa
with
\beqa \label{covder2}
D_\mu Q^{\rm em}_L &=& \partial_\mu Q^{\rm em}_L 
- i[l_\mu,Q^{\rm em}_L], \nn
D_\mu Q^{\rm em}_R &=& \partial_\mu Q^{\rm em}_R - i[r_\mu,Q^{\rm em}_R].
\eeqa
In $\cL_{\rm lept}$  we consider only terms quadratic in the lepton
fields and at most linear in $G_F$. The terms with $X_{4,5}$ will not appear in
realistic physical processes as the generated amplitudes
contain an external (axial-) vector source (see Eqs. (\ref{covder1}) 
and (\ref{covder2})).
  
In deriving a minimal set of terms in Eq. (\ref{Llept}), we have
used partial integration, the equations of motion derived from
the tree-level Lagrangian (\ref{Leff})
and the relations
\beq
\Q_L^{\rm em} \Q_L^{\rm w} = \frac{2}{3} \Q_L^{\rm w} , \qquad  
\Q_L^{\rm w} \Q_L^{\rm em}  = -\frac{1}{3} \Q_L^{\rm w} , \qquad 
\langle \Q_L^{\rm w}  \rangle = 0.
\eeq

Finally, also a photon Lagrangian
\beq
\cL_{\gamma} = e^2 X_8 F_{\mu \nu} F^{\mu \nu}, \qquad 
F_{\mu \nu} = \partial_{\mu} A_{\nu} - \partial_{\nu} A_{\mu},
\eeq
has to be added. This term cancels the divergences of the photon
two-point function generated by the lepton loops.

The ``new'' low--energy couplings $X_i$ arising here are
divergent (except $X_1$). In the dimensional regularization
scheme, they absorb the divergences of the
one--loop graphs with internal lepton lines via the renormalization
\beqa
X_i &=& X_i^r(\mu) + \Xi_i \Lambda(\mu) , \quad i=1,\ldots,8 , \no \\
\Lambda(\mu) &=& \frac{\mu^{d-4}}{(4\pi)^2} \left\{ \frac{1}{d-4} -
\frac{1}{2} [\ln (4\pi) + \Gamma'(1) + 1]\right\}. \label{renorm}
\eeqa 

The coefficients $\Xi_1, \ldots \Xi_7$  can be 
determined \cite{KNRT99} by using super-heat-kernel methods 
\cite{Berezinian,SHK}:
\beqa
&& \Xi_1 = 0, \quad \Xi_2 = -\dfrac{3}{4}, \quad 
\Xi_3 = -3, \quad \Xi_4 = -\dfrac{3}{2}, 
 \nn [5pt]   
&& \Xi_5 = \dfrac{3}{2}, \quad \Xi_6 = -5, 
\quad \Xi_7 = -1, \quad \Xi_8 = -\dfrac{4}{3}.
\eeqa

\paragraph*{4.}

We have developed the appropriate low-energy
effective theory for a complete treatment of isospin violating
effects in semileptonic weak processes. The electromagnetic
interaction requires the inclusion of the photon field and the
light leptons as explicit dynamical degrees of freedom in the
chiral Lagrangian. At next-to-leading order, the list of local
terms given by Gasser and Leutwyler \cite{GL85} for the QCD part
and by Urech \cite{Urech} for the electromagnetic interaction of
the pseudoscalars has to be enlarged. This is, of course, a
consequence of the presence of virtual leptons in our extended
theory. Regarding pure lepton or photon bilinears as
``trivial'', five additional ``non-trivial'' terms of this type
are arising. But two of them will not appear in realistic
physical processes. One may therefore conclude that the main
bulk of electromagnetic low-energy constants is already
contained in Urech's Lagrangian and the inclusion of virtual
leptons in chiral perturbation theory does not substantially aggravate
the problem of unknown parameters.

As an illustration of the use of our effective theory, we have calculated
\cite{KNRT99} the decay rates of $\pi \ra \ell \nu_{\ell}$ and
$K \ra \ell \nu_{\ell}$ including the electromagnetic contributions of
$\cO (e^2 p^2)$. An investigation of the 
$K_{\ell 3}$ 
decays is
presently in progress.

The continuation of our work will follow two principal
lines. Firstly, we are now in the position to calculate the
electromagnetic contributions to $K_{\ell 3}$ and $K_{\ell 4}$ 
decays where
all constraints imposed by chiral symmetry are taken into
account. In spite of our large ignorance of the actual values 
of the electromagnetic low-energy couplings, it will often be
possible to relate the electromagnetic contributions to
different processes. For specific combinations of
observables one might even find parameter-free predictions.
Simple examples of this kind have been given for the $P_{\ell
2}$ decays \cite{KNRT99}. In some fortunate cases simple
order-of-magnitude
estimates for the electromagnetic couplings based on chiral 
dimensional analysis may even be sufficient.

Secondly, a further major task for the next future is, of course, the
determination of the physical values of the electromagnetic
low-energy coupling constants 
in the standard model. In contrast to the
QCD low-energy couplings $L_1^r, \cdots L_{10}^r$ which are rather well
determined from experimental input and large $N_c$ arguments within the
standard framework of chiral perturbation theory \cite{GL85}, only very
little is known so far in the electromagnetic sector. First attempts to
estimate some of the Urech constants $K_i$ can be found in 
\cite{BB96,BB97,Moussallam}. As far as the constants $X_i$ are concerned,
the recent analysis \cite{KPPdR99} of the counterterms contributing to the
decay processes of light neutral pseudoscalars into charged lepton pairs
raises hopes that reliable estimates for these constants can be achieved
within a large-$N_c$ approach.

\end{document}